# An Analysis of Blockchain Adoption in Supply Chains Between 2010 and 2020

**Nikhil Vadgama[1], Paolo Tasca**

Centre for Blockchain Technologies
University College London
(United Kingdom)

**Abstract**

In this research, the evolution of Distributed Ledger Technology (DLT) in supply chains has been mapped from the inception of the technology until June 2020, utilising primarily public data sources. Two hundred seventy-one blockchain projects operating in the supply chain have been analysed on parameters such as their inception dates, types of blockchain, stages reached, sectors applied to and type of organisation that founded the project. We confirm generally understood trends in the blockchain market with the creation of projects following the general hype and funding levels in the industry. We observe most activity in the Agriculture/Grocery sector and the Freight/Logistics sector. We see the shift of market interest from primarily private companies (startups) to public companies and consortia and the change in blockchain adoption from Ethereum to Hyperledger. Finally, we observe higher success and lower failure rates for Hyperledger-based projects in comparison to Ethereum-based projects.

---

## 1    Introduction

Distributed Ledger Technology (DLT) promises to disrupt business models, business processes, and aspects of society by creating information systems that are transparent and provide a single point of truth for all members of a network (Pilkington, 2016). As an electronic ledger that has the properties of decentralisation, immutability, cryptography and smart contracts, DLT represents an innovation beyond traditional database technology (Iansiti & Lakhani, 2017).

Since the creation of DLT with its beginnings with Bitcoin in 2008 (Nakamoto, 2008), DLT's impact has begun to move outside of just the financial services domain into other sectors, including that of supply chains (Bünger, 2017; Hughes, Park, Archer-Brown, & Kietzmann, 2019). Within the supply chain domain, it is widely acknowledged by many industry experts that DLT will have a tremendous impact on it, particularly around bringing transparency across various parts of it (O'Marah, 2017; Casey & Wong, 2017).

Supply chains underpin the smooth and timely movement of goods from producer through to the consumer, and with increasing globalisation, this coordination of goods underpins the globalised economy. The supply chain management sector stands at a size of $16 trillion and has large overhead with regards to fraud, errors and administration costs (Boucher, 2017).

---

[1] Correspondence to Nikhil Vadgama: nikhil.vadgama@ucl.ac.uk

Given the importance of this part of the world economy, it is astonishing that there is still a large degree of manual procedures and processes in operationally complex undertakings. For example, the shipment of refrigerated goods between East Africa and Europe can incorporate as many as 30 different individuals and organisations and involve over 200 different interactions and communications. The cost of processing all the paperwork associated with a shipment can easily be around 15% of the shipment costs (Groenfeldt, 2017).

In light of recent events with the Covid-19 pandemic, the fragility of our current supply chains and globalised trade operations were exposed (Lin & Lanng, 2020). In particular, the importance of bringing supply chain infrastructure up to speed through digitisation is important, and emerging technologies will play a part as the enabling force for economic, business and social transformation (Morkunas, Paschen & Boon, 2019).

The relevance of blockchain for supply chains has already been widely discussed in academic literature (Kshetri, 2018). Much of the literature focuses on Bitcoin and explores potential applications. It does not describe the state of the market and its evolution (Min, 2019). Some studies exist that focus on survey-based methods including Petersen, Hackius, and von See (2018) who survey supply chain professionals on the use of blockchain and Wamba, Queiroz and Trinchera (2020) who survey practitioners to investigate the drivers of blockchain adoption in the supply chain.[2]

This study offers a different perspective by exploring the state of blockchain adoption in supply chains based on publicly available data. As part of the research for this paper, 271 relevant blockchain projects were analysed in the supply chain domain. Analysis of the data gathered through this research supports the narrative of both the general trends observed in the blockchain supply chain domain and concerning project inception dates, types of blockchain utilised, stages projects reached, sectors applied to and type of organisation that founded the project. We confirm generally understood trends in the blockchain market with the creation of projects following the general hype and cryptocurrency market prices and funding levels in the market. We observe most activity happening in the Agriculture/Grocery sector and the Freight/Logistics sector. We see the shift in market interest from primarily private companies (startups) to public companies and consortia and the change in blockchain adoption from Ethereum to Hyperledger and the success and failure rates of projects that adopt these blockchains.

## 2    Materials and Methods

Information on 271 projects utilising blockchain for supply chain purposes was collected and analysed between the period of May and June 2020. Information was found from a variety of different sources through desktop research primarily of company websites, news articles, and data repositories. Project information found was between 2010 and the first half of 2020. Raw data for 31 different fields were collected. An explanation of some of the more relevant fields for data collection is presented below, and in Appendix 1 (Table 5) and a full example of the cleaned data collected is presented for one particular project in Appendix 2 (Table 6).

The methodological approach taken was first to collect raw data from multiple sources and then to clean it. Thereafter, descriptive statistics were utilised to generate information on overall trends in the data. Finally, inferential statistics were applied to generate insights. Analysis was performed on the parameters of time, stage, blockchain, sector, and organisation type.

There were numerous difficulties in the collection of the dataset. This included the variety of sources that were required in order to gather information, as often data from one source would point to another with each presenting some new information that would be relevant for a particular project. There are also issues on the veracity of the data and whether the information found was completely verifiable. Where possible, every effort was taken to verify claims made by projects and companies through manual cross-validation. This aspect is discussed further below and in the challenges and the limitations of the research section. With regards to the main fields of analysis and data collected, this is further elucidated below.

---

[2] In this article we use the term 'blockchain' to refer also to the larger family of distributed ledger technologies (DLT), i.e., community consensus-based distributed ledgers where the storage of data is not based on chains of blocks.

**Project Organisation Classification:** Projects were classified based on the type of organisation that was leading a particular project. Rather than define a project by an organisation, the term project is used as a single organisation could have many different projects, and each project was classified separately. The four types of organisations that were used to classify projects were:

- *Private companies* - labelled as startups as the vast majority were early-stage companies.
- *Public companies* - were those that are listed on public equity markets.
- *Consortia* - were identified by having either a separate legal entity or website and a degree of separation from the individual organisations taking part in the project (i.e. not being an organisation and its clients).
- *Government project* - if a government body led a project.

**Stage of development**: several different classifications were used to describe the stage a project was at including:

- *Failed* – explicit confirmation was found that the project was abandoned, social network accounts were no longer active, and the website had been taken down or that the project had no update for longer than one year.
- *Development* – the project is in the ideation stage.
- *Pilot* – the project has moved from ideation to trial and proof-of-concept.
- *Production* – the project is successful and being deployed in industry and available for partners and clients to purchase and use.

**Sectors**: Sectors that were considered for classification of the data were firstly based on the SIC classification system. This was too broad, and so a sector classification based on the natural and obvious sectoral classification of the projects was adopted. Sectors were classified if at least 1% of the data was present; otherwise, these projects were placed in the category *Other*. A *Multiple* classification was used for projects that were not sector-specific. Clearly, identifiable sectors in the data that emerged were:

- *Aerospace and Defence*
- *Agriculture / Grocery*
- *Automotive*
- *Fashion*
- *Finance*
- *Freight / Logistics*
- *Luxury items*
- *Mining*
- *Oil & Gas*
- *Pharma / healthcare*

- *Multiple*
- *Other*

**Blockchains** - The major blockchains classified include those that were identifiable in at least 1% of the data; otherwise, they were classified as *Other*. In many projects, the blockchain was not disclosed, and these projects had their blockchain categorised as *TBC*. Where a blockchain was utilising an existing codebase and was not completely distinct from it, the blockchain from which the codebase derived was used to classify the project. The group *Agnostic* comprises solutions that were not tied to any particular blockchain. The blockchains classifications used were:

- *Ant Blockchain*
- *Bitcoin*
- *Corda*
- *Ethereum*
- *Hyperledger*
- *Oracle Blockchain*
- *Quorum*
- *VeChain*

- *Agnostic*
- *Other*
- *TBC*

## 3    Results

We present the results on 271 blockchain projects with respect to the year in which the project was created, the blockchains used, the stage the project has reached, the organisation leading the project and the sector the project was applied in.

Figure 1 shows the number of projects with respect to their founding year. The peak of projects being created is in 2018, with 56.8% of all projects founded in 2017 and 2018 alone. After 2018 we see the number of projects fall. 2020's data is only partial for the year (until June), but already has nearly as many projects as 2015. No projects were discovered that were founded in 2011.

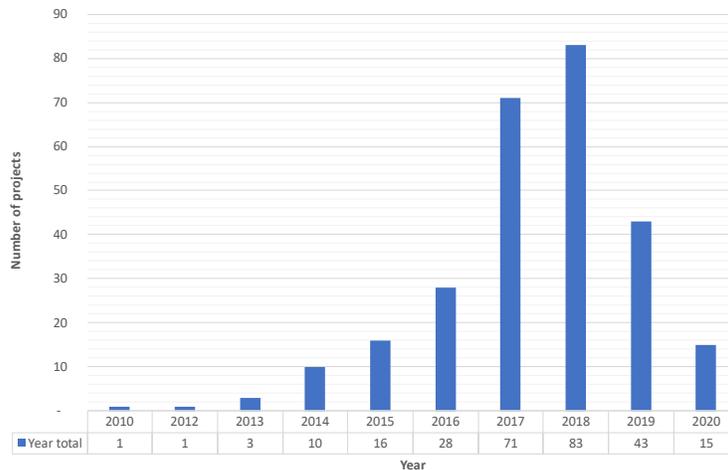

***Figure 1:*** *The number of blockchain projects created in each year*

Figure 2 shows the percentage of projects that use different blockchains based on their founding years. The major blockchains adopted are Ethereum, utilised by 22.9% of all projects and Hyperledger, utilised by 21.4% of all projects. 12.5% of projects are blockchain agnostic.[3] 22.9% of projects do not disclose the blockchain that they use and are in the *TBC* category. Either these projects are still experimenting or deciding which blockchains to use (and in some cases, these projects no longer exist)[4], or they are operating and have not disclosed what type of blockchain solution they utilise.[5] We also see that projects developed on the Ethereum platform were more prevalent than Hyperledger projects in 2015, 2016, and 2017, whilst this is opposite for projects created in 2018, 2019 and for 2020 so far.

---

[3] An example project in this category would be Origintrail (https://origintrail.io), who have a protocol that can work with different blockchain solutions.
[4] See for example Resonance (https://www.digicatapult.org.uk/for-startups/success-stories/resonance)

The greatest proportion of Ethereum projects were from companies created in 2017, with 40.3% of all Ethereum based projects created in this year. This is approximately one to two years after the release of Ethereum in July 2015 (Ethereum, 2020), indicating a lag in the creation of projects using this blockchain (as it appears today). On the other hand, the greatest proportion of Hyperledger projects was in 2018 with 37.9% of all projects utilising Hyperledger. Again this also follows a lag of one to two years after the creation of Hyperledger in late 2015, early 2016 (Hyperledger, 2020). Projects in the Agnostic group accounted for the largest percentage of projects founded in 2014, but fluctuate under 20% over other years. Finally, projects in the TBC group are approximately 25-30% of all projects in each of 2016, 2017 and 2018.

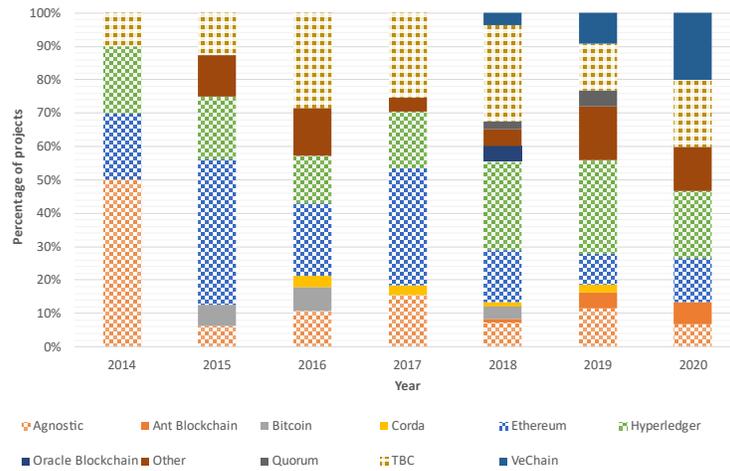

*Figure 2: The percentage of projects using a particular blockchain based on the project's year of creation*

Figure 3 shows the stages that projects have reached based on the year they were created. Out of the entire dataset, 23.2% of projects reached the production stage, 45.4% of projects were in pilot, 21.8% in development and 9.6% were identified as failed. The greatest proportion of projects in production are from 2014, 2015, and 2017 with 50%, 37.5% and 31% of projects respectively. 2016 appears to be an anomaly with only 14.3% of projects from that year reaching production. The greatest proportion of projects in production occurred in 2017. 34.9% of all production projects were created in that year. Many projects also appear stuck in the pilot and development stages, and a minority of projects have also failed. 2017 and 2018 feature the greatest proportion of failed projects, with 34.6% and 46.2% of all failed projects (80.8%) occurring from projects created in these years.

---

[5] See for example Remedichain https://www.remedichain.org

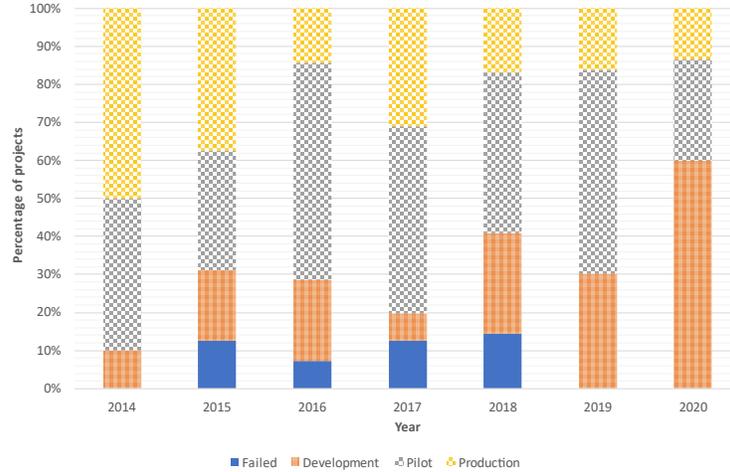

*Figure 3: The percentage of projects at various stages based on the project's year of creation*

Figure 4 shows the sectors that projects operate in based on their founding year. The top three dominant sectors are Agriculture/Grocery, Freight/Logistics and Multiple sectors at 39.5%, 17% and 12.5% of all projects. Agriculture/Grocery projects account for over 40% of all projects founded in 2014, 2015, 2017, 2018 and 2019. The second most popular sector is Freight/Logistics, which reaches a peak of 35.7% of all projects created in 2016. Finally, 31.8% of all Agriculture/Grocery projects were in 2018 alone, whilst 30.4% of all Freight/Logistics projects were in 2017 alone.

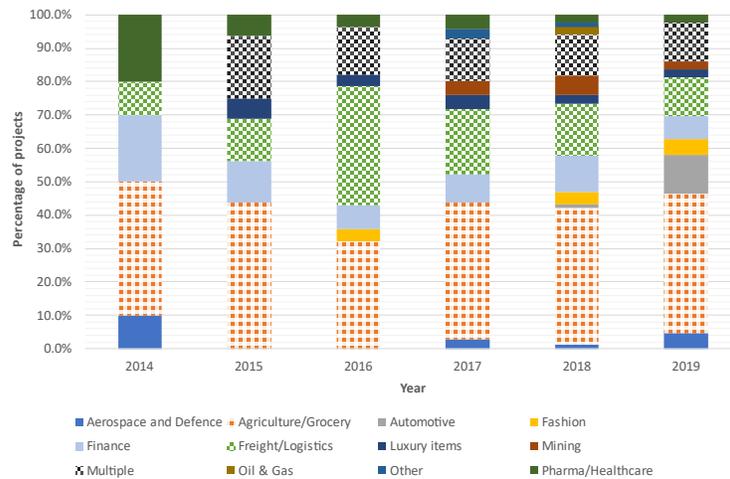

*Figure 4: The percentage of projects operating in sectors based on the project's year of creation*

Figure 5 shows the lead organisation of a project by the founding year of the project. Startups (private companies) account for 64.2% of the entire dataset, followed by public companies at 17.3%, consortium at 14.8% and finally, government initiatives at 3.7%. Startups account for all projects in 2014 and 2015 and then decline over time, accounting for only 25.6% of all projects created in 2019. Other types of organisations enter into the fray from 2016 onwards. Of all startup projects, 34.5% of them were created in 2017 alone.

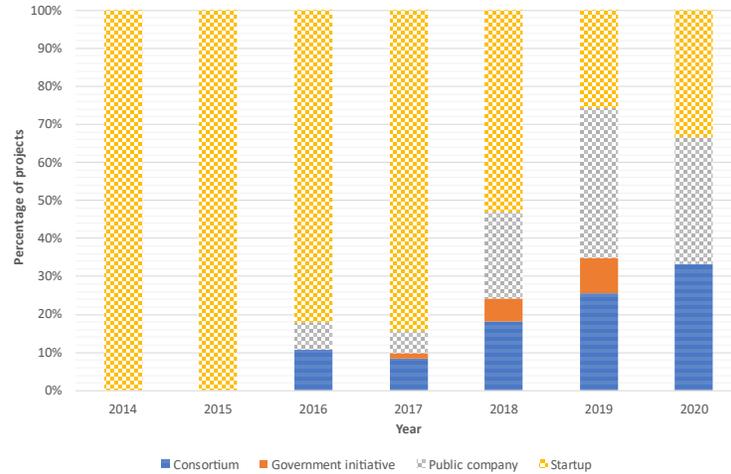

***Figure 5:** The percentage of different types of organisation leading projects based on the project's year of creation*

After analysing projects along various dimensions based on the year that they were created, we now analyse all the projects irrespective of time. Figure 6 shows the stages that projects reached based on the applied blockchain solution. Of interest is that Hyperledger projects tend to be more successful on the whole, with 33.3% of projects at the production stage utilising this blockchain. Also of interest is that of those projects identified as having failed, 34.6% of these were utilising the Ethereum platform. Other interesting points to note are that many projects have reached advanced stages without revealing the blockchain technologies that they are working with, which is for example why 16% of projects at the production stage are in the *TBC* category. Of all the projects that utilise Hyperledger, 46.6% of them have reached the pilot phase, 36.2% of them the production phase and only 1.7% have failed. In comparison, to all Ethereum projects, 50% have reached the pilot stage, 21% the production stage and 14.5% have failed.

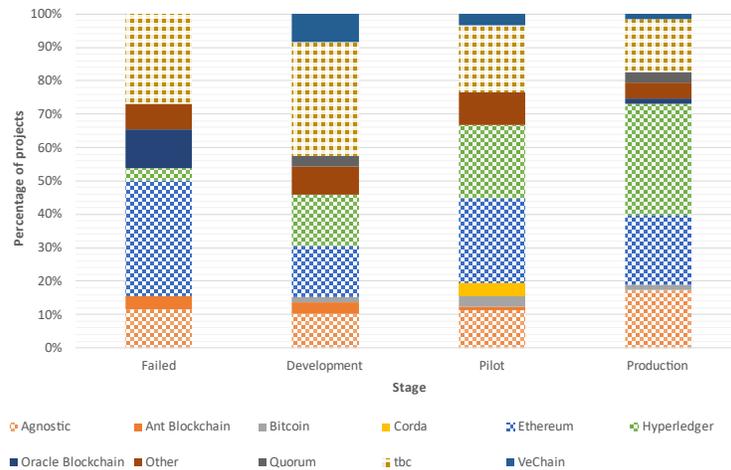

**Figure 6:** *The percentage of projects using a particular blockchain based on the stages projects reach*

Figure 7 shows the sectors in which the projects are applied in. Sectors which have at least 5% of all projects in the dataset are the Agriculture/Grocery, Freight/Logistics, Multiple and Finance sectors. Looking at the Agriculture/Grocery sector (with 39.5% of projects overall), 23.4% of all the Agriculture/Grocery projects utilised Ethereum, 18.7% Hyperledger, 8.4% *Agnostic*, 8.4% *Other* and 33.6% were *TBC*. In comparison, for the Freight/Logistics sector (17% of projects overall), 28.3% utilised Ethereum, 10.9% Hyperledger, 13% *Agnostic* and 26.1% *TBC*. For the *Multiple* sectors category, Hyperledger was most popular with 35.3% of projects utilising this blockchain, followed by 23.5% of projects that were *Agnostic* and 8.8% utilising Ethereum. Within the Finance sector (9.2% of projects overall), 24% of projects were in the *TBC* category, 20% utilised Ethereum, and 12% *Agnostic*, Hyperledger and Corda each. Out of all Ethereum-based projects, 40.3% were focused on the Agriculture sector, and 21% in the Freight/Logistics sector. For Hyperledger, 34.5% of projects were in the Agriculture/Grocery sector and 8.6% in the Freight/Logistics sectors. For those projects that did not disclose the blockchains they are working with, 58.1% are focussed on Agriculture.

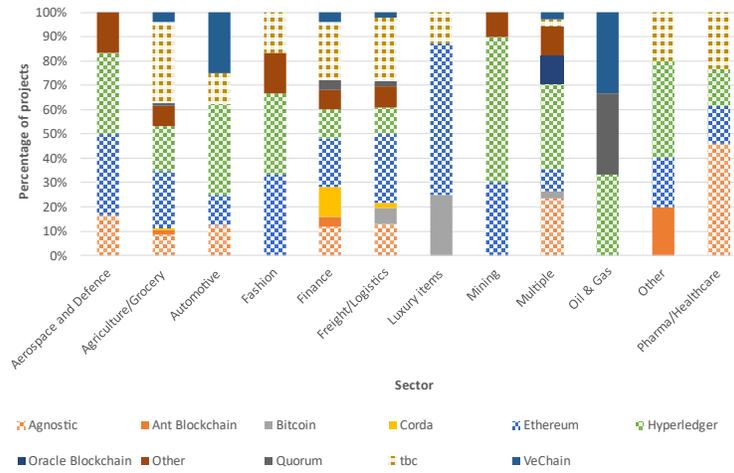

***Figure 7:*** *The percentage of projects using a particular blockchain based on the sector a project operates in*

Figure 8 shows the blockchains used by projects based on the type of lead organisation. Ethereum dominates the startup category, with 32.2% of all startup projects utilising this platform. For consortia and public companies, Hyperledger is the most popular used 35% and 34% respectively. For government initiatives, 50% of projects did not identify the blockchains that they utilised (*TBC*). Out of all Ethereum projects, 90% were utilised by startups. Of all Hyperledger projects, consortia, public companies and startups utilised this blockchain at 24.1%, 27.6% and 44.8%. Overall, startups utilising Ethereum accounted for 20.7% of the dataset as the largest single group.

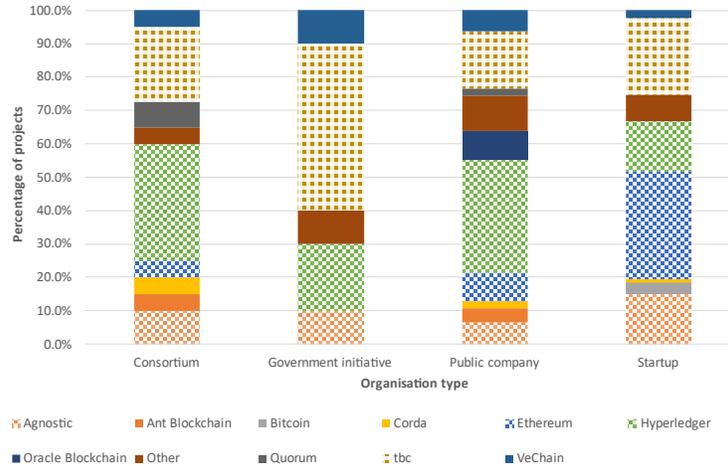

***Figure 8:** The percentage of projects using a particular blockchain based on the type of organisation leading the project*

Figure 9 shows the stages reached by different projects based on the type of organisation leading them. Startups - accounting for 64.2% of the entire dataset - have the greatest degree of success in reaching the production stage, with 28.7% of them reaching this stage. Out of all public company projects, 17% of them have reached the production stage, and for government initiatives and consortia, 10% reach this stage each. With regards to failure, startups have the lowest failure rate at 8.6%, followed by consortia and public companies with 10% and 10.6% respectively in their categories. Government initiatives have the highest failure rate at 20% of all projects failing.

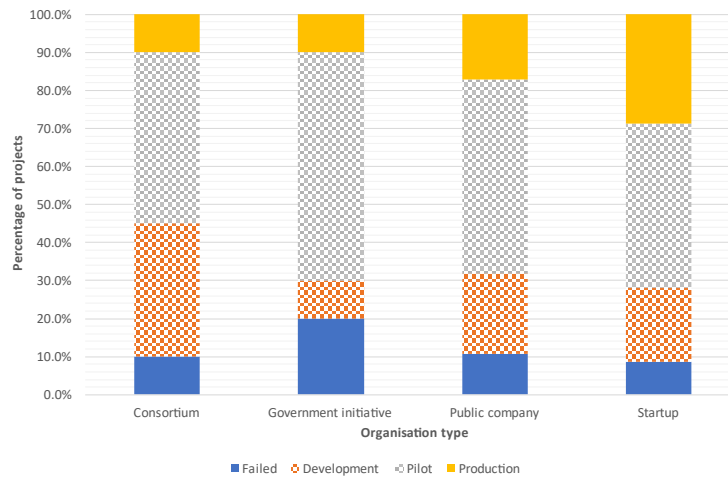

***Figure 9:** The percentage of projects at various stages based on the type of organisation leading the project*

Figure 10 shows the projects that operate in different sectors based on the type of organisation leading the project. Agriculture/Grocery projects account for the greatest share in each of the different organisation types with 25%, 60% 31.9% and 43.7% of consortia, government initiatives, public company and startups respectively. Freight/Logistics is the second-largest sector that projects operate in with 22.5%, 20%, 4.3% and 19% of projects in each of the consortia, government initiative, public company and startup categories. Startups are also engaged in the most number of sectors, followed by consortia and public companies. Government initiatives operate in the least number of sectors.

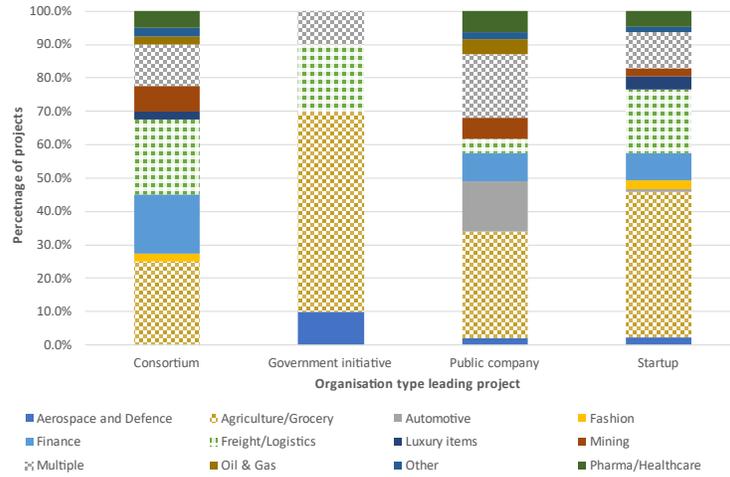

***Figure 10:** The percentage of projects operating in different sectors based on the type of organisation leading the project*

Figure 11 shows the projects at various stages within different sectors. Starting with the most popular sectors of Agriculture/Grocery, Freight/Logistics and Multiple, the percentage of projects within these sectors reaching production are 22.4%, 21.7% and 29.4%, respectively. The sectors with the highest percentage of projects in production are Aerospace and Defense at 33.3%, Oil and Gas at 33.3%, Other at 40% and Pharma/Healthcare at 30.8%. However, it should be noted that the number of projects in these sectors is between 2-5% of the entire dataset. Out of those projects that failed, we observe a failure rate of 14.7% in the *Multiple* category, 10.9% for the Freight/Logistics sector and 8.4% for the Agriculture/Grocery sector.

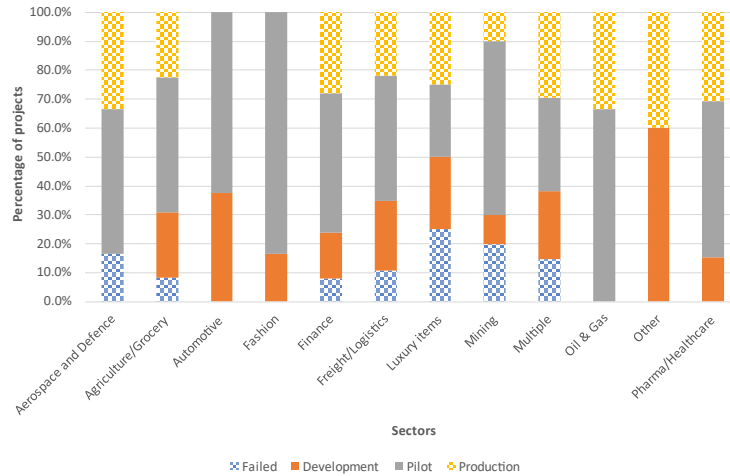

***Figure 11:** The percentage of projects at the stages they have reached within different sectors*

In this last part of the analysis, we look more closely at projects that are utilising the Hyperledger and Ethereum blockchains and also at projects operating in the Agriculture/Grocery and Freight/Logistics sectors. We look at these subsets as these two sectors, and blockchains account for 56.5% and 44.3% of all projects respectively.

Focusing on Ethereum based projects only, Table 1 shows the stages reached by the projects broken down into their leading organisations. Here we see that the vast majority of Ethereum projects are led by startups (90.3%) compared to 64.2% of projects in our entire dataset. With respect to stages, the Ethereum projects failure rate (14.5%) is greater than the entire dataset (9.6%). Finally of all Ethereum projects, 21% reach the production stage compared with 23.2% from the entire dataset.

| Organisation Type | Stage (% of all Ethereum based projects) | | | | | Organisation total (for all projects) |
|---|---|---|---|---|---|---|
| | Failed | Development | Pilot | Production | Organisation total | |
| **Consortium** | 0.0% | 1.6% | 1.6% | 0.0% | 3.2% | **14.8%** |
| **Government Initiative** | 0.0% | 0.0% | 0.0% | 0.0% | 0.0% | **3.7%** |
| **Public company** | 0.0% | 1.6% | 4.8% | 0.0% | 6.5% | **17.3%** |
| **Startup** | 14.5% | 11.3% | 43.5% | 21.0% | 90.3% | **64.2%** |
| **Stage total (% of all Ethereum based projects)** | 14.5% | 14.5% | 50.0% | 21.0% | | |
| **Stage total (% of all projects)** | **9.6%** | **21.8%** | **45.4%** | **23.2%** | | |

***Table 1:*** *Ethereum based projects and the stages that they have reached based on the leading organisation of the project*

We repeat the same analysis as above for Hyperledger-based projects only. Table 2 shows the stages reached by the Hyperledger projects broken down into their leading organisations. Differently from Ethereum projects, here we see that the minority of Hyperledger projects are led by startups (44.8%). We also observe that Hyperledger projects are led by consortia and public companies 24.1% and 28%, respectively. This is higher than the dataset average, where we find that consortia and public companies led for 14.8% and 17.3% of all projects respectively. Moreover, we find that of all Hyperledger projects, 36.2% reach the production stage. This percentage is higher than the average production rate of all projects (23.2%). Also, Hyperledger projects' failure rate is very low (1.7%) compared with the average failure rate of all projects (9.6%).

| Organisation Type | Stage (% of all Hyperledger based projects) | | | | | Organisational total (for all projects) |
|---|---|---|---|---|---|---|
| | Failed | Development | Pilot | Production | Organisation total | |
| **Consortium** | 1.7% | 1.7% | 17.2% | 3.4% | 24.1% | **14.8%** |
| **Government Initiative** | 0.0% | 0.0% | 3.4% | 0.0% | 3.4% | **3.7%** |
| **Public company** | 0.0% | 5.2% | 13.8% | 8.6% | 27.6% | **17.3%** |
| **Startup** | 0.0% | 8.6% | 12.1% | 24.1% | 44.8% | **64.2%** |
| **Stage total (% of all Hyperledger based projects)** | 1.7% | 15.5% | 46.6% | 36.2% | | |
| **Stage total (% of all projects)** | **9.6%** | **21.8%** | **45.4%** | **23.2%** | | |

***Table 2:*** *Hyperledger based projects and the stages that they have reached based on the leading organisation of the project*

Table 3 compares Hyperledger and Ethereum projects in the Agriculture/Grocery and Freight/Logistics sectors. Here we see that Hyperledger has a greater percentage of projects in production and less failed projects than Ethereum in the Agriculture/Grocery sector. For the Freight/Logistics sector, Ethereum has more production and failed projects than Hyperledger. To summarise, Hyperledger based projects tend to perform better by showing a lower failure rate and a higher production rate not only when compared with Ethereum based projects but also for the entire dataset as well.

| Blockchain | Sector / Stage | **Failed** | **Development** | **Pilot** | **Production** |
|---|---|---|---|---|---|
| **Ethereum (% of all Ethereum projects in that sector or stage)** | Agriculture/Grocery | 12.0% | 12.0% | 60.0% | 16.0% |
| | Freight/Logistics | 15.4% | 15.4% | 23.1% | 46.2% |
| | **Stage total** | **14.5%** | **14.5%** | **50.0%** | **21.0%** |
| **Hyperledger (% of all Hyperledger projects in that sector or stage)** | Agriculture/Grocery | 0.0% | 20.0% | 45.0% | 35.0% |
| | Freight/Logistics | 0.0% | 0.0% | 80.0% | 20.0% |
| | **Stage total** | **1.7%** | **15.5%** | **46.6%** | **36.2%** |
| **Percentage of all projects at each stage** | | **9.6%** | **21.8%** | **45.4%** | **23.2%** |

***Table 3:*** *Comparison of projects that use Hyperledger and Ethereum within the Agriculture/Grocery and Freight/Logistics sectors*

Table 4 shows the percentage of projects using either Hyperledger or Ethereum in a particular year and within a particular sector in comparison with the averages for all sectors. Here we can see that for the Agriculture/Grocery sector, Ethereum usage is higher for projects created in 2014, 2015, and 2016 than Ethereum-based projects in all sectors and for Hyperledger projects. For 2017-2020, the proportion of Ethereum projects is lower than Ethereum projects in all sectors. For 2018-2020, the proportion of Hyperledger projects are greater than Ethereum in the Agriculture/Grocery sector. For the Freight/Logistics sector, the proportion of Ethereum projects tend to be in greater than Hyperledger (except in 2014 where they are equal and 2018 where there are more Hyperledger projects).

| Sector | Blockchain / Year | 2014 | 2015 | 2016 | 2017 | 2018 | 2019 | 2020 |
|--------|-------------------|------|------|------|------|------|------|------|
| **Agriculture (% of projects in the sector that use that particular blockchain)** | Ethereum | 50.0% | 71.4% | 22.2% | 27.6% | 14.7% | 5.6% | 0.0% |
| | Hyperledger | 25.0% | 28.6% | 22.2% | 13.8% | 17.6% | 22.2% | 50.0% |
| **Freight (% of projects in the sector that use that particular blockchain)** | Ethereum | 0.0% | 100.0% | 20.0% | 50.0% | 7.7% | 20.0% | 0.0% |
| | Hyperledger | 0.0% | 0.0% | 10.0% | 7.1% | 23.1% | 0.0% | 0.0% |
| **All sectors (blockchain used as a percentage of all projects from that year)** | Ethereum | 20.0% | 43.8% | 21.4% | 35.2% | 15.7% | 9.3% | 13.3% |
| | Hyperledger | 20.0% | 18.8% | 14.3% | 16.9% | 26.5% | 27.9% | 20.0% |

***Table 4:*** *Comparison of Ethereum and Hyperledger projects in the Agriculture and Freight and Logistics sectors over time*

## 4      Discussion

Based on the results, a number of general findings emerge that showcase how blockchain projects are evolving in the supply chain arena. Many of the findings fit industry reporting, news and survey results that are available in the public domain (for example, see Deloitte, 2020). In line with general interest indicators such as google trends (search term "blockchain") (Google, 2020) and cryptocurrency prices (Bitcoin's peak price in late 2017) (CoinMarketCap, 2020), we these patterns of peak interest and price match the pattern of new projects being created. New project formation peaks in 2018 and then drops off in 2019 and 2020. This is most clearly seen in Figure 1, showcasing the number of projects by their year of creation.

As well as the number of projects being created each year, we see the shift in the market from being a startup, and private company dominated to enterprise and consortium dominated. With regards to startups and private companies, as shown in Figure 5, we see the percentage of private company projects trending from 100% in 2014 and gradually decreasing to 25.6% in 2019 and 33.3% for the first half of 2020. This shows that more public companies are coming to the fray to engage in projects, as well as consortia and government initiatives. Public companies account for the majority at 39.5% of all projects in 2019.

Further confirming the movement from private companies to public company led projects is the change in the blockchain platforms utilised. Of identifiable blockchains, Ethereum represents 22.9% of projects in the sample, and Hyperledger represents 21.4% of projects. We see that Ethereum projects account for 43.8%, 21.4% and 35.2% of projects created in 2015, 2016, 2017, and then Hyperledger dominating in 2018, 2019, 2020 with 26.5%, 27.9% and 20% respectively as seen in Figure 2. Part of this switch is, of course, the difference in the inception of the different blockchains. Ethereum was launched in 2015 (Ethereum, 2020), whilst Hyperledger in 2016 (Hyperledger, 2020). The use of these blockchains can be seen in Figure 8. As public companies become dominant in 2018 and 2019, we see that 34% of all public company projects utilise Hyperledger, whilst the 32.2% of all startup projects utilise Ethereum.

One interesting fact point to examine is the number of projects that do not identify the blockchain that they use. As shown in Figure 2, overall, 22.9% of all the projects are in the *TBC* category. Either this is because the projects are still in ideation, or pilot phases where they wish to keep this information proprietary. 72.6% of the projects in the *TBC* category are in the development and pilot stages, more than for Ethereum or Hyperledger (64.5% and 62.1% of those projects). *TBC* projects account for the largest percentage of all projects that are in the development stage at 33.9%, which represents the largest grouping and suggests that these projects are still more in the ideation stage.

Concerning stages, one would expect that projects that have existed for longer (and created earlier) would perhaps have a greater chance of experimenting and getting to the production stage. This is confirmed in Figure 3 where we see that projects created in 2014 and 2015 reached the production stage at 50% and 37.5%, respectively. With respect to failures, 2018 has the highest failure rate, with 14.5% of all projects created in that year having failed. Due to the large number of projects created on the back of hype in the sector, this may indicate these projects were less thought through and therefore had lower chances of success. In particular, the boom in Initial Coin Offerings (ICO) occurred with the peak in cryptocurrency prices and the formation of projects, decreasing the barrier to starting a project as funding was more accessible. Funding peaked for ICO's in 2018 at $12.62 billion (Liu, 2020), after the peak in Bitcoin's price in late 2017 (CoinMarketCap, 2020). The pattern of projects created we observe very much fits the funding cycle for 2017, 2018 and 2019 (Tasca,Vigliotti & Gong, 2018).

Another interesting fact to note is the relatively large number of projects that remain in the development and pilot stages. 67.2% of all projects are in the development and pilot stages, with 27.7% of all projects at these stages founded in the years between 2014 and 2017. There may be two reasons for this. The first could be that projects are particularly complex and take many years to move through stages. This is certainly described in the literature (see, e.g. Iansiti and Lakhani, 2017) which states that it will take years for blockchain as a foundational technology to change the supply chain landscape. The second reason could be that organisations wish to signal that projects are still alive to either not admit failure, or to be ready to revive a project when the time in the market for deployment is right. For projects led by public companies, consortia, and government initiatives, a greater percentage than the dataset average (67.2%) are in the development and pilot stages.

With respect to stages and organisations, we see from Figure 9 that the greatest proportion (28.7%) of startup projects are in the production stage. Government projects had the highest failure rate at 20% of all government projects. This may be due to the fact that government projects may have a greater degree of complexity than private sector projects due to legislative issues and accompanying bureaucracy and this may explain the reason fewer government initiatives have succeeded.

With respect to sectors, as shown in Figure 4, Agriculture/Grocery dominates throughout all years and accounts for 39.5% of the dataset. Food safety is of paramount importance, and so is the ability to track and trace agriculture and grocery products. This may explain why the majority of projects that utilise blockchain are concentrated in this sector. The need for this is brought particularly to light given scandals in recent years where there have been incidents of milk powder contamination (Xiu & Klein, 2010), E-coli outbreaks (Casey & Wong, 2017) and meat substitution (Falkheimer & Heide, 2015).

Freight/Logistics is the second-largest sector accounting for 17% of all projects. As discussed earlier, the complexity of moving products for example from East Africa to Europe required over 200 interactions and involved more than 30 individuals and organisations in the journey and had costs of paperwork exceeding 15% of the entire transportation cost (Bapai, 2017 & Groenfeldt, 2017). This complexity implies there are potentially large efficiency gains that could be made which explains the attention given to these sectors. Agriculture is also the dominant sector amongst organisation types leading projects and is largest for government-led projects, with 60% of these projects taking place in the agriculture sector (above the 39.5% for all projects). Out of all consortia led projects, only 22.5% are in the Freight/Logistics sector, greater than the dataset average (14.8%). This reflects the relative coordination complexity in the sector, implying greater coordination of stakeholders needed, which consortia with their governance structures may be able to facilitate.

Of most interest is the comparison between Ethereum and Hyperledger projects; and the comparison between projects in Agriculture/Grocery and Freight/Logistics sectors.

The results show that Ethereum has a much greater proportion of startups (90% of all projects) than the average across all projects (Table 1). This is very different from Hyperledger, which has a lower proportion in startups (44.8% of all projects), but a relatively higher proportion of consortia and public companies (Table 2). This distinction can be attributed to the different nature of the blockchains. Ethereum is a public blockchain and relatively easy to fundraise for, particularly during the ICO boom as seen in 2017. Hyperledger, on the other hand, is more suited for private usage and therefore fits use by enterprises, or public companies.

On the proportion of projects that have failed compared to those that reached production, Ethereum shows a higher degree of failure and a lower level of production than the average rates of all the projects in the dataset. On the opposite side, Hyperledger projects show a lower failure rate than the average rate of all the projects in the dataset. Moreover, Hyperledger projects show a higher production rate. It is interesting to note why this may be the case, and perhaps this is to do with the nature of blockchain implementation. It is much easier to implement projects amongst an ecosystem if one is the dominant player. Public companies are much more likely to exhibit this behaviour, and certainly, we see this in the case of Hyperledger with 24% of all production projects coming from public companies and 31% of all public companies utilising Hyperledger reaching this stage.

The above results may find an explanation on Wamba, Queiroz and Trinchera's (2020) findings according to which knowledge sharing and trading partner pressure lead to successful outcomes for the adoption of blockchain in supply chains. This supports the assertion that public companies (due to their size and influence) are more likely to be able to create pressure on organisations in their ecosystem to adopt blockchain, thereby potentially leading to greater success for projects. Indeed, the very small number of public companies, consortia or government initiatives leading Ethereum based projects can explain the lower production stage statistics we see for this blockchain.

Finally, Table 4 shows the trends in Hyperledger and Ethereum adoption over time for all projects and the Agriculture/Grocery and Freight/Logistics sectors. As discussed earlier, the shift from Ethereum to Hyperleger can be seen occurring in 2018. This pattern is also seen in the Agriculture/Grocery sector, but not in the Freight/Logistics sector. This may be explained by the fact that Freight/Logistics projects have greater complexity and involve cross border provenance, for instance, requiring the use of public blockchains over private ones, and hence the utilisation of Ethereum here. Indeed, we can observe that Ethereum outperforms Hyperledger in the Freight/Logistics sector because 46% of those projects reached production compared to 20% for Hyperledger.

## 5    Challenges in the research, limitations and future directions

This research's limitations are primarily based on the difficulty of sourcing full information on the nature of blockchain projects in supply chains. Although there exist some repositories of project information, there are still many more projects that were found by searching online, looking at company websites and examining general press and news reports. It was also difficult to find good quality data on the nature of blockchains in supply chains. This limited the amount of analysis that could be completed. For example, it would be interesting to examine the funding levels for projects. However, within our sample, funding data was only available for 28% of projects and therefore deemed not large enough to draw meaningful conclusions.

Our analysis also focussed on projects with information accessible in English. This precluded many projects that are assumed to be occurring in China. This inference can be made by looking at patent applications. China has accounted for nearly 60% of the total number of blockchain applications submitted by the USA, China, Japan, South Korea and Germany altogether through 2018, with its application total being nearly three times larger than the USA (Chen, 2020). Given this, the fact that only 7.7% of projects in our dataset were operating in China would indicate Chinese projects were most likely underrepresented. This is most likely as Chinese project information is not published in English and therefore, could not be included in this study.

Many projects may not accurately represent their true status with, for example, the large number of projects in the development and pilot stages. Some projects have been at these stages for many years. It is interesting to speculate why they are still active on communications and on their websites (as mentioned in the discussion above). This may undoubtedly be explained as there is a reason to keep a project going on for marketing purposes in case their use case will be useful in the future. Production-level projects may also be overestimated as organisations can simply state they have production-level projects without actually having any other stakeholders utilising their solutions. For example, a startup that builds a blockchain solution that is ready to use can be classified as production-ready, even though there may not be any users.

This research is also a snapshot of the state of the blockchain market historically from today's perspective. This means that although some elements of the market's evolution have been presented, the full extent of all trends in the industry cannot be analysed. For example, some projects in 2013 and 2014 are using the Ethereum blockchain. This is today's snapshot of their behaviour as Ethereun was not launched until the middle of 2015. It was not possible to see what solutions they were using before Ethereum and, when and why they switched technologies.

Finally, it would be interesting to extend the dimensions of this research into other variables if enough information could be found on funding levels, application areas in the supply chain and even to examine if the current pandemic situation has led to more opportunities for implementing DLT based solutions. Furthermore, this research has not examined in detail reasons for why projects have succeeded or failed beyond looking at the statistics and making inferences. It would be interesting to explore several cases with interviews. This may paint a better picture of the market's evolution of blockchain usage over time and enable discussion of best practises leading to more successful project outcomes for the deployment of blockchain in supply chains.

## 6    Conclusion

In this research, we have begun to map out how blockchain has evolved with respect to its usage in the supply chain sector. Utilising a number of different parameters, we have investigated which sectors have seen projects take place, which blockchains are utilised, what organisations are leading and how successful projects have been. We have observed that the greatest concentration of projects is in the Agriculture/Grocery and Freight/Logistics sectors. We have confirmed market trends that blockchain projects have shifted from being startup (private company) led to public company led and that the most popular blockchain used has changed from Ethereum to Hyperledger. Finally, we see that Hyperledger based projects have a greater success rate and lower failure rate than Ethereum concerning our entire dataset.

# 7    Appendix 1

| Field Name | Explanation of Field |
| --- | --- |
| Project name | The name of the project, or company if only a single company was leading this project |
| Website | The website of the project |
| Type of Organisation behind the Project | The type of project, whether this was:<br>• Startup (or private company)<br>• Government initiative<br>• Public Company<br>• Consortium |
| Sector of operation | Assessment of most suitable sector for the project |
| Project location | What country(s) the project is primarily operating in |
| Region | What region(s) the project is primarily operating in |
| Year of founding | The year the project was founded |
| Stage of project | Stage the project has reached:<br>• Failed<br>• Development<br>• Pilot<br>• Production |
| Organisations Involved in the Project | Other organisations involved in the project if any |
| Name of DLT utilised | The DLT that was primarily utilised |

***Table 5:*** *Explanation of the main fields of data collection.*

## 8        Appendix 2

| Field | Information |
|---|---|
| Project name | Ambrosus |
| Website | [https://ambrosus.com/](https://ambrosus.com/) |
| Type of Organisation behind the Project | Startup |
| Sector of operation | Pharma / healthcare |
| Project location | Switzerland |
| Region | EMEA |
| Year of founding | 2017 |
| Stage of project | Pilot |
| Organisations Involved in the Project | Nongshim, UN 10YFP, European Institute of Innovation & Technology |
| Name of DLT utilised | AMB-NET (Ethereum) |

**Table 6:** *Example of project-specific fields.*